\documentclass[10pt]{article}
\usepackage{amsmath}
\usepackage{amsfonts}

\usepackage{mathrsfs,amsthm,amsmath,amssymb}
\usepackage{graphicx}
\usepackage{color}
\usepackage{url}

\theoremstyle{definition}

\newtheorem{Lemma}{Lemma}

\newtheorem{Conjecture}{Conjecture}

\newcommand{\fpm}{{\mathbb F}_{p^m}}
\newcommand{\fq}{{\mathbb F}_{q}}
\newcommand{\fqtwo}{{\mathbb F}_{q^2}}

\newcommand{\Proof}{\noindent\textbf{Proof.}~}
\newcommand{\done}{\hfill $\Box$ }

\newcommand{\ls}[1]
    {\dimen0=\fontdimen6\the\font\lineskip=#1\dimen0
     \advance\lineskip.5\fontdimen5\the\font
     \advance\lineskip-\dimen0
     \lineskiplimit=0.9\lineskip
     \baselineskip=\lineskip
     \advance\baselineskip\dimen0
     \normallineskip\lineskip\normallineskiplimit\lineskiplimit
     \normalbaselineskip\baselineskip
     \ignorespaces}

\begin{document}

\bibliographystyle{abbrv}


\title{On Two Conjectures about Permutation Trinomials over  $\mathbb{F}_{3^{2k}}$}
\author{ Nian Li 
\thanks{Department of Informatics, University of Bergen,
 N-5020 Bergen, Norway. Email: Nian.Li@uib.no}
}
\date{}
\maketitle
\ls{1.5}

\thispagestyle{plain} \setcounter{page}{1}

\begin{abstract}
 Permutation polynomials with few terms attracts researchers' interest in recent years due to their simple algebraic form and some additional extraordinary properties. In this paper, by analyzing the quadratic factors of a fifth-degree polynomial and a seventh-degree polynomial over the finite field $\mathbb{F}_{3^{2k}}$, two conjectures on permutation trinomials over $\mathbb{F}_{3^{2k}}$ proposed recently by Li, Qu, Li and Fu are settled, where $k$ is a positive integer.
\end{abstract}

\section{Introduction}

A permutation polynomial over a finite field is a polynomial that acts as a permutation of the elements of the filed. Permutation polynomials were first studied by Hermite \cite{Hermite} for prime fields and by Dickson \cite{Dickson} for arbitrary finite fields. Permutation polynomials over finite fields  have wide applications in coding theory, cryptography and combinatorial designs, and it is of great interest in both theoretical and practical aspects to find new permutation polynomials.  The reader is referred to \cite{BTT,Ding-QWYY,Hou14,Hou,Hou15,Li-Qu-Chen,Tu-Zeng-Hu,Tu-Zeng-Hu-Li,Tu-Zeng-J,Tu-Zeng-L-H,Zeng,ZTT,ZZC,Zieve-subgroup} for some constructions of permutation polynomials over finite fields.

Let $p$ be a prime, $m$ be a positive integer and $\fpm$ denote the finite field with $p^m$ elements. An important class of permutation polynomials over the finite field $\fpm$ is of the form
\begin{eqnarray}\label{form}
x^rh(x^{\frac{p^m-1}{d}}),
\end{eqnarray}
where $r,d$ are positive integers satisfying $d\,|\,p^m-1$, $1\leq r<(p^m-1)/d$ and $h(x)\in\fpm[x]$. This class of permutation polynomials originated from the work of Dickson \cite{Dickson-58}, Carlitz and Wells \cite{Carlitz-Wells} and Niederreiter and Robinson \cite{Niederreiter-Robinson} who considered some special cases of the form \eqref{form}. Wan and Lidl  \cite{Wan-Lidl} first provided a unified criterion for a polynomial with the form \eqref{form} to be a permutation polynomial in terms of the primitive roots.  Later, Park and Lee \cite{Park-Lee} and Zieve \cite{Zieve-09} studied this class of permutation polynomials and showed that the polynomial with the form \eqref{form} is a permutation polynomial if and only if $\gcd(r,(p^m-1)/d)=1$ and $x^rh(x)^{(p^m-1)/d}$ permutes the set $\mu_d$ of the $d$-th  roots of unity in $\fpm$. In this sense, to determine the permutation property of \eqref{form}, the crucial step is to decide whether $x^rh(x)^{(p^m-1)/d}$ permutes $\mu_d$. However, this is still a difficult problem in general.

The construction of permutation trinomials with the form \eqref{form} reattracts researchers' attention due to a recent work of Ding et al \cite{Ding-QWYY}. Motivated by Ding et al.'s work, some new permutation trinomials of the form \eqref{form} were obtained in \cite{GS,Li-Qu-Chen,LQLF,LH} by using different approaches in solving equations with low degree over finite fields and in \cite{LH2} by using the property of the linear fractional polynomials over finite fields. The results obtained in \cite{LH2} are the generalizations of some works in  \cite{Ding-QWYY,GS,Li-Qu-Chen,LQLF}. In a very recent paper \cite{LQLF}, Li et al. constructed several classes of permutation trinomials of the form \eqref{form} with $m=2k$ and $d=p^k+1$ for $p=2, 3$ and proposed three conjectures on permutation trinomials in such form for $p=3$. This paper is devoted to settle two of the conjectures proposed in \cite{LQLF}. The key step to solve the conjectures is to prove that a fifth-degree equation and a seventh-degree equation over $\mathbb{F}_{3^{2k}}$ have a unique solution in $\mu_{3^k+1}$, i.e., the set of the $(3^k+1)$-th  roots of unity in $\mathbb{F}_{3^{2k}}$. By analyzing the quadratic factors of the correseponding fifth-degree and seventh-degree polynomials over $\mathbb{F}_{3^{2k}}$, we can obtain their possible quadratic factors and then the conjectures can be settled based on further discussions on the solutions to the quadratic factors.

The remainder of this paper is organized as follows. Section \ref{sec-conj} introduces some notations and the conjectures proposed in \cite{LQLF}. Sections \ref{sec-pc-3} and \ref{sec-pc-12} prove two of the conjectures by analyzing the quadratic factors of a fifth-degree polynomial and a seventh-degree polynomial over $\mathbb{F}_{3^{2k}}$ respectively, and some concluding remarks are given in Section \ref{sec-cr}.

\section{The two conjectures on permutation trinomials over $\mathbb{F}_{3^{2k}}$}\label{sec-conj}

Let $p$ be a prime and $m$ be a positive integer. A criterion for a polynomial in the form \eqref{form} to be a permutation polynomial had been characterized by the following lemma which was proved by Park and Lee in 2001 and reproved by Zieve in 2009. 

 \begin{Lemma}(\cite{Park-Lee,Zieve-09}) \label{lem-Z}
  The polynomial defined as in \eqref{form} is a permutation over $\fpm$ if and only if
 \begin{enumerate}
   \item [(1)] $\gcd(r,(p^m-1)/d)=1$, and
   \item [(2)] $x^rh(x)^{(p^m-1)/d}$ permutes the set of the $d$-th roots of unity in $\fpm$.
 \end{enumerate}
\end{Lemma}

Lemma \ref{lem-Z} reduces the problem of determination of permutations over $\fpm$ to that of determination of permutations over its subgroups. However, it is still a difficult problem to verify the second condition in Lemma \ref{lem-Z}. By using Lemma \ref{lem-Z}, Li, Qu, Li and Fu presented several classes of permutation trinomials of the form \eqref{form} with $m=2k$ and $d=p^k+1$ for $p=2, 3$ by solving certain low-degree equations over finite fields and finally they proposed three conjectures on permutation trinomials over $\mathbb{F}_{3^{2k}}$.

 From now on, let $m=2k$ be a positive integer and $q=3^k$. The set of the $(q+1)$-th  roots of unity in $\fqtwo$ is given as follows:
\[\mu_{q+1}=\{x\in\fqtwo:  x^{q+1}=1\}.\]

 \begin{Conjecture}(Conjecture 5.1, \cite{LQLF}) \label{conj-1} \vspace{-2mm}
 \begin{enumerate}
   \item [(1)] Let $q=3^k$, $k$ be even and $f(x)=x^{lq+l+5}+x^{(l+5)q+l}-x^{(l-1)q+l+6}$, where $\gcd(5+2l,q-1)=1$. Then $f(x)$ is a permutation trinomial over $\fqtwo$.  \vspace{-2mm}
   \item [(2)] Let $q=3^k$, $f(x)=x^{lq+l+1}-x^{(l+4)q+l-3}+x^{(l-2)q+l+3}$ and $\gcd(1+2l,q-1)=1$. Then $f(x)$ is a permutation trinomial over $\fqtwo$. \vspace{-2mm}
   \item [(3)] Let $q=3^k$, $f(x)=x^{lq+l+1}+x^{(l+2)q+l-1}-x^{(l-2)q+l+3}$ and $\gcd(1+2l,q-1)=1$. Then $f(x)$ is a permutation trinomial over $\fqtwo$ if $k\not\equiv 2\pmod{4}$. \vspace{-2mm}
 \end{enumerate}
\end{Conjecture}

According to Lemma \ref{lem-Z},  Conjecture \ref{conj-1} is equivalent to the following conjecture.

 \begin{Conjecture}(Conjecture 5.2, \cite{LQLF}) \label{conj-2} \vspace{-2mm}
 \begin{enumerate}
   \item [(1)] Let $q=3^k$, $k$ be even and $g(x)=\frac{-x^7+x^6+x}{x^6+x-1}$. Then $g(x)$ permutes $\mu_{q+1}$.  \vspace{-2mm}
   \item [(2)] Let $q=3^k$ and $g(x)=\frac{x^6+x^4-1}{-x^7+x^3+x}$. Then $g(x)$ permutes $\mu_{q+1}$. \vspace{-2mm}
   \item [(3)] Let $q=3^k$ and $g(x)=\frac{-x^5+x^3+x}{x^4+x^2-1}$. Then $g(x)$ permutes $\mu_{q+1}$ if $k\not\equiv 2\pmod{4}$. \vspace{-2mm}
 \end{enumerate}
\end{Conjecture}

To prove Conjecture \ref{conj-2}, we need to show that for any $g(x)$ listed above the equation $g(x)=t$ has a unique solution in $\mu_{q+1}$ for any $t\in\mu_{q+1}$. Normally, it is a hard problem to determine the number of solutions to an equation (even with low degree) over finite fields. As pointed out by the authors in \cite{LQLF}, the main difficulty to prove these conjectures lies in dealing with some specified equations with high degree. In what follows,  we aim to settle Conjecture \ref{conj-2} (2) and Conjecture \ref{conj-2} (3), for this goal we determine the quadratic factors of a fifth-degree polynomial and a seventh-degree polynomial and then show that  $g(x)=t$ cannot have distinct solutions in $\mu_{q+1}$ respectively.

\section{Proof of Conjecture \ref{conj-2} (3)}\label{sec-pc-3}

To prove Conjecture \ref{conj-2} (3), we first show that $x^4+x^2-1=0$ has no solution in $\mu_{q+1}$. Otherwise, we have $x^{q+1}=1=x^4+x^2$ which implies that $x^q=x^3+x$. Taking $q$-th power on both sides gives $x=x^{3q}+x^q=x^{-3}+x^{-1}$ which leads to $x^4-x^2-1=0$, a contradiction with $x^4+x^2-1=0$ and $x\in\mu_{q+1}$. On the other hand, it can be readily verified that $g(x)^{q+1}=1$ for any $x \in \mu_{q+1}$. Thus, to prove Conjecture \ref{conj-2} (3), it suffices to show that $\frac{-x^5+x^3+x}{x^4+x^2-1}=t$ has a unique solution in  $\mu_{q+1}$ for any $t\in\mu_{q+1}$ if $k\not\equiv 2\pmod{4}$, which is equivalent to proving that the equation
\begin{eqnarray}\label{eq-5th}
 x^5+tx^4-x^3+tx^2-x-t=0
\end{eqnarray}
has at most one solution in $\mu_{q+1}$  for any $t\in\mu_{q+1}$ if $k\not\equiv 2\pmod{4}$.

\begin{Lemma}\label{lem-5th-ab}
Let $t\in\mu_{q+1}$ and $F(x)=x^5+tx^4-x^3+tx^2-x-t$. If $x^2+ax+b$, where $a,b\ne 0$, is a quadratic factor of $F(x)$, then $a,b$ must satisfy 
$a^2=(\epsilon-1)b^2-(\epsilon+1)b+\epsilon-1$, where $\epsilon^2+1=0$.
\end{Lemma}

\Proof Assume that $F(x)$ can be factorized as $F(x)=(x^2+ax+b)(x^3+\sigma_1x^2+\sigma_2x+\sigma_3)$. Expanding the right hand side of $F(x)$ and comparing the coefficients of $x^{4-i}$ for $i=0,1,\cdots,4$ gives
\begin{eqnarray} \label{eq-coe}
a+\sigma_1=t, \;b+\sigma_2+a\sigma_1=-1, \;b\sigma_1+a\sigma_2+\sigma_3=t, \;a\sigma_3+b\sigma_2=-1, \;b\sigma_3=-t.
\end{eqnarray}
From the first two identities in \eqref{eq-coe} we have $b+\sigma_2+a(t-a)=-1$ and then $\sigma_2=a^2-at-b-1$, and by the last two identities in \eqref{eq-coe} we have $a(-t/b)+b\sigma_2=-1$ which leads to $\sigma_2=(at-b)/b^2$. This implies that $a^2-at-b-1=(at-b)/b^2$, i.e., 
\begin{eqnarray}\label{eq-5th-1}
(a+ab^2)t=a^2b^2-b^3-b^2+b.
\end{eqnarray}
On the other hand, by the third identity in \eqref{eq-coe} we have $b^2\sigma_1+ab\sigma_2+b\sigma_3=bt$ which implies that $b^2(t-a)+a(-1-a\sigma_3)+(-t)=bt$. Replacing $\sigma_3$ by $(-t/b)$ gives $b^3t-b^3a-ab+a^2t-bt=b^2t$, then we obtain that 
\begin{eqnarray}\label{eq-5th-2}
(b^3-b^2-b+a^2)t=ab^3+ab.
\end{eqnarray}

We then can discuss the relation between $a$ and $b$ as follows:

Case 1:  $a+ab^2=0$. If this case happens, then we have $b^2=-1$ since $a,b\ne 0$ and then \eqref{eq-5th-1} implies $a^2b^2-b^3-b^2+b=0$ which leads to $a^2b^2-b(b^2-1)-b^2=0$, i.e., $a^2=1-b$. By $b^2=-1$ we have $ab^3+ab=0$ and then $b^3-b^2-b+a^2=0$ due to \eqref{eq-5th-2}, i.e., $a^2=b^2+b(1-b^2)=-1-b$, a contradiction with $a^2=1-b$. Thus, this case cannot happen.

Case 2:  $b^3-b^2-b+a^2=0$. For this case, we then have $ab^3+ab=0$ according to \eqref{eq-5th-2}, i.e., $b^2=-1$ since $a,b\ne 0$ and then from $b^3-b^2-b+a^2=0$ we have $a^2=b^2+b(1-b^2)=-1-b$. Again by $b^2=-1$ we can obtain $a+ab^2=0$ which leads to $a^2b^2-b^3-b^2+b=0$ by \eqref{eq-5th-1}, i.e., $a^2=1-b$, a contradiction with $a^2=-1-b$. Thus, this case cannot happen either.

Case 3: $a+ab^2\not=0$ and $b^3-b^2-b+a^2\not=0$. In this case, by \eqref{eq-5th-1} and \eqref{eq-5th-2} we have
\[\frac{a^2b^2-b^3-b^2+b}{a+ab^2}=\frac{ab^3+ab}{b^3-b^2-b+a^2}\]
which is equivalent to
\begin{eqnarray*}
a^4-(b-1)^2a^2-(b^4+1)=0
\end{eqnarray*}
since $ab\ne 0$. Note that the discriminant of the above quadratic equation on variable $a^2$ is  $\Delta=(b-1)^4+4(b^4+1)=-(b+1)^4$. This implies
\[a^2=\frac{(b-1)^2\pm\epsilon(b+1)^2}{2}=(-1\pm\epsilon)b^2+(-1\mp\epsilon)b+(-1\pm\epsilon),\]
where $\epsilon^2+1=0$.  Then the result follows from $\epsilon^2+1=(-\epsilon)^2+1=0$. This completes the proof. \done

\begin{Lemma}\label{lem-5th}
For any $t\in\mu_{q+1}$ \eqref{eq-5th} cannot have distinct solutions in $\mu_{q+1}$ if $k\not\equiv 2\pmod{4}$.
\end{Lemma}

\Proof Suppose that \eqref{eq-5th} have two distinct solutions $x_1, x_2\in\mu_{q+1}$, this means that the polynomial $x^5+tx^4-x^3+tx^2-x-t$ has a quadratic factor $x^2+ax+b$ satisfying $x_1+x_2=-a$ and $x_1x_2=b$. Moreover, the two solutions can be expressed as
\begin{eqnarray*}
 x_1=a-\sqrt{a^2-b},\; x_2=a+\sqrt{a^2-b}.
\end{eqnarray*}
This together with Lemma \ref{lem-5th-ab} implies that
\begin{eqnarray} \label{eq-5th-x12}
 x_1=a-\sqrt{\epsilon-1}(b+1),\; x_2=a+\sqrt{\epsilon-1}(b+1),
\end{eqnarray}
where $\epsilon^2+1=0$. Note that \eqref{eq-5th} has repeated roots if $b=-1$ and in this case the repeated root is $x=\epsilon$ according to Lemma \ref{lem-5th-ab}.  Next let $b\ne -1$  and $\alpha$ be a primitive element of $\fqtwo$, then $\epsilon=\pm\alpha^{(q^2-1)/4}$ and by $\epsilon^2+1=0$ we have $(\epsilon-1)^2=\epsilon$. Then we can discussion \eqref{eq-5th-x12} as below:
\begin{enumerate}  \vspace{-2mm}
  \item [(1)] $k\equiv 1, 3\pmod{4}$. For this case, we can obtain that $q^2-1\equiv 8\pmod{16}$ which implies that $\sqrt{\epsilon-1}\not\in\fqtwo$ due to $(\epsilon-1)^2=\epsilon$. Thus \eqref{eq-5th} cannot have two distinct solutions in $\mu_{q+1}$ if $k\equiv 1, 3\pmod{4}$.  \vspace{-2mm}
  \item [(2)] $k\equiv 0\pmod{4}$. In this case, by $x_1, x_2\in\mu_{q+1}$, we have $(a\pm\sqrt{\epsilon-1}(b+1))^{q+1}=1$ which leads to
  \begin{eqnarray} \label{eq-5th-s}
  a^q\sqrt{\epsilon-1}(b+1)+a\sqrt{\epsilon-1}^q(b^q+1)=0.
  \end{eqnarray}
On the other hand, according to $x_1+x_2=-a$ and $x_1x_2=b$, we have $b\in\mu_{q+1}$ and $-a^q=x_1^q+x_2^q=(x_1+x_2)/(x_1x_2)=-a/b$, i.e., $b^q=b^{-1}$ and $a^q=a/b$. Therefore, \eqref{eq-5th-s} can be reduced to $\sqrt{\epsilon-1}+\sqrt{\epsilon-1}^q=0$ since $a,b,(b+1)\ne 0$. This is impossible since $\sqrt{\epsilon-1}^{q-1}=(\sqrt{\epsilon-1}^4)^{(q-1)/4}=\epsilon^{(q-1)/4}=(-1)^{(q-1)/8}=1$ due to $(\epsilon-1)^2=\epsilon$ and $q-1\equiv 0\pmod{16}$.  \vspace{-2mm}
\end{enumerate}
Combining the above cases, we can conclude that \eqref{eq-5th} cannot have distinct solutions in $\mu_{q+1}$ if $k\not\equiv 2\pmod{4}$. This completes the proof. \done

According to Lemma \ref{lem-5th}, we complete the proof of Conjecture \ref{conj-2} (3). Next we prove Conjecture \ref{conj-2} (2) in the same manner by analyzing the quadratic factors of a seventh-degree polynomial over $\fqtwo$.

\section{Proof of Conjecture \ref{conj-2} (2) }\label{sec-pc-12}

Notice that $-x^7+x^3+x\ne 0$ if $x \in \mu_{q+1}$. Otherwise we have $x^6-x^2=1=x^{q+1}$ and then $x^q=x^5-x$.  Taking $q$-th power on both sides gives $x=x^{5q}-x^q=x^{-5}-x^{-1}$ which leads to $x^6+x^4-1=0$. This together with $x^6-x^2-1=0$ gives $x^4+x^2=0$, i.e., $x^2=-1$, a contradiction. Thus, we have $-x^7+x^3+x\ne 0$ if $x \in \mu_{q+1}$. Then, for any $x\in\mu_{q+1}$, we have $g(x)^{q+1}=1$, i.e., $g(x)\in\mu_{q+1}$. Therefore, to complete the proof of Conjecture \ref{conj-2} (2), it is sufficient to show that $=\frac{x^6+x^4-1}{-x^7+x^3+x}=\frac{1}{t}$ has a unique solution for any $t\in\mu_{q+1}$, i.e., the equation
\begin{eqnarray}\label{eq-7th}
 x^7+tx^6+tx^4-x^3-x-t=0
\end{eqnarray}
has at most one solution in $\mu_{q+1}$  for any $t\in\mu_{q+1}$ for any positive integer $k$.

Similar as the proof of Conjecture \ref{conj-2} (3), we next show that  \eqref{eq-7th} cannot have distinct solutions in $\mu_{q+1}$ for any $t\in\mu_{q+1}$ by analyzing the possible quadratic factors of $ x^7+tx^6+tx^4-x^3-x-t$ over $\fqtwo$. Note that if  \eqref{eq-7th}  have distinct solutions in $\mu_{q+1}$ then $ x^7+tx^6+tx^4-x^3-x-t$ must have a quadratic factor $x^2+ax+b$ for some $a,b\in\fqtwo$ satisfying $a^qb=a$ according to the proof of Lemma \ref{lem-5th}.

\begin{Lemma}\label{lem-7th-ab}
Let $t\in\mu_{q+1}$ and $G(x)=x^7+tx^6+tx^4-x^3-x-t$. If $x^2+ax+b$, where $a,b\ne 0$ and $a^qb=a$, is a quadratic factor of $G(x)$, then $a,b$ must satisfy one of the following conditions:
\begin{enumerate}  \vspace{-2mm}
  \item [(1)] $(a,b)=(\epsilon, -1)$, where $\epsilon^2+1=0$. \vspace{-2mm}
  \item [(2)] $a^2=\theta  b^2-(\theta-1)b+\theta$, where $\theta^3-\theta-1=0$. \vspace{-2mm}
\end{enumerate}
\end{Lemma}

\Proof Assume that $G(x)$ can be factorized as $G(x)=(x^2+ax+b)(x^5+\sigma_1x^4+\sigma_2x^3+\sigma_3x^2+\sigma_4x+\sigma_5)$. Expanding the right hand side of $G(x)$ and comparing the coefficients of $x^{6-i}$ for $i=0,1,\cdots,6$ gives
\begin{eqnarray} \label{eq-coe1-7th}
a+\sigma_1=t, \;b+a\sigma_1+\sigma_2=0, \;b\sigma_1+a\sigma_2+\sigma_3=t, \;b\sigma_2+a\sigma_3+\sigma_4=-1
\end{eqnarray}
and 
 \begin{eqnarray} \label{eq-coe2-7th}
b\sigma_3+a\sigma_4+\sigma_5=0, \;a\sigma_5+b\sigma_4=-1, \;b\sigma_5=-t.
\end{eqnarray}
By a direct calculation, from \eqref{eq-coe1-7th} we have $\sigma_1=t-a$, $\sigma_2=-a\sigma_1-b=a^2-at-b$, $\sigma_3=t-a\sigma_2-b\sigma_1=t-a^3+a^2t-bt-ab$ and $\sigma_4=-1-a\sigma_3-b\sigma_2=-1-at+a^4-a^3t-abt+b^2$.  On the other hand, by \eqref{eq-coe2-7th}, we can obtain $\sigma_5=-t/b$, $\sigma_4=(-1-a\sigma_5)/b=(at-b)/b^2$ and $\sigma_3=(-\sigma_5-a\sigma_4)/b=(ab+bt-a^2t)/b^3$. Therefore, by the value of $\sigma_3$ we get $t-a^3+a^2t-bt-ab=(ab+bt-a^2t)/b^3$, which can be written as
\begin{eqnarray}\label{eq-7th-ab1}
 (a^2b^3+a^2-b^4+b^3-b)t=a^3b^3+ab^4+ab,
\end{eqnarray}
and according to the value of $\sigma_4$ we have $-1-at+a^4-a^3t-abt+b^2=(at-b)/b^2$,  i.e.,
\begin{eqnarray}\label{eq-7th-ab2}
 (a+ab^2+a^3b^2+ab^3)t=a^4b^2+b^4-b^2+b.
\end{eqnarray}

Then we can discuss \eqref{eq-7th-ab1} and \eqref{eq-7th-ab2} as follows:

Case 1:  $a^2b^3+a^2-b^4+b^3-b=0$.  For this case, by \eqref{eq-7th-ab1} we have $a^3b^3+ab^4+ab=0$, i.e., $a^2b^2+b^3+1=0$ since $ab\ne 0$. Replacing $b$ by $a/a^q$ and then multiplying by $a^{3q}$ gives $a^4a^q+(a+a^q)^3=0$, which leads to $a\in\fq$ due to $aa^q, (a+a^q)\in\fq$. Hence, again by $a^qb=a$ we get $b=1$ and then $a^2=1$. This contracts with $a^2b^3+a^2-b^4+b^3-b=0$. Thus this case cannot occur. 

Case 2:  $a+ab^2+a^3b^2+ab^3=0$, i.e., $a^2=-(b^3+b^2+1)/b^2$.  Then, \eqref{eq-7th-ab2} implies that $a^4b^2+b^4-b^2+b=0$ which gives $a^4=-(b^3-b+1)/b$. From these two identities one gets $(b^3+b^2+1)^2/b^4=-(b^3-b+1)/b$, which can be reduced to $b^6+b^5+b^2-1=0$. Note that $b^6+b^5+b^2-1=(b^6-1)+b^2(b^3+1)=(b^3+1)(b^3+b^2-1)$. Thus we get $b=-1$ or $b^3+b^2-1=0$. If $b=-1$, then $a^2=-(b^3+b^2+1)/b^2=-1$. If $b^3+b^2-1=0$, then $a^2=-(b^3+b^2+1)/b^2=1/b^2$, which means $a^2\in\mu_{q+1}$ since $b=a^{1-q}\in\mu_{q+1}$, i.e., $a^{2q+2}=1$. Moreover, substituting $b=a^{1-q}\in\mu_{q+1}$ into $a^2b^2=1$ gives $a^{4-2q}=1$, which leads to $a^{6-2q-2}=a^6=(a^2)^3=1$. That is, $a^2=1$ and then $b^2=1$, a contradiction with  $b^3+b^2-1=0$ due to $b\ne 0$. Hence, this case happens only if $a^2=-1$ and $b=-1$.

Case 3:  $a^2b^3+a^2-b^4+b^3-b\not=0$ and $a+ab^2+a^3b^2+ab^3\not=0$.  By \eqref{eq-7th-ab1} and \eqref{eq-7th-ab2},  we have
\[\frac{a^3b^3+ab^4+ab}{a^2b^3+a^2-b^4+b^3-b}=\frac{a^4b^2+b^4-b^2+b}{a+ab^2+a^3b^2+ab^3}\]
which can be reduced to $a^6-(b+1)^4a^2-(b^6-b^5-b^4-b^2-b+1)=0$ by a straight calculation since $ab\ne 0$. If $b=-1$, then we have $a^6+1=(a^2+1)^3=0$, i.e., $a^2=-1$. If  $b\not=-1$, then the above equation can be rewriten as
\begin{eqnarray*}
 (\frac{a^2}{(b+1)^2})^3-\frac{a^2}{(b+1)^2}-\frac{b^6-b^5-b^4-b^2-b+1}{(b+1)^6}=0.
\end{eqnarray*}
Observe that $\frac{b^6-b^5-b^4-b^2-b+1}{(b+1)^6}=\frac{b^6+1}{(b+1)^6}-\frac{b(b+1)^4}{(b+1)^6}=\frac{b^6+1}{(b+1)^6}-\frac{b}{(b+1)^2}=1+\frac{b^3}{(b+1)^6}-\frac{b}{(b+1)^2}$. Then, the above equation can be further written as
\begin{eqnarray*}
 (\frac{a^2}{(b+1)^2}-\frac{b}{(b+1)^2})^3-(\frac{a^2}{(b+1)^2}-\frac{b}{(b+1)^2})-1=0,
\end{eqnarray*}
which implies that $\frac{a^2}{(b+1)^2}-\frac{b}{(b+1)^2}=\theta$, i.e., $a^2=\theta  b^2-(\theta-1)b+\theta$, where $\theta^3-\theta-1=0$. 
This completes the proof. \done

According to Lemma \ref{lem-7th-ab}, we can obtain the following desired result.

\begin{Lemma}\label{lem-7th}
For any $t\in\mu_{q+1}$ \eqref{eq-7th} cannot have distinct solutions in $\mu_{q+1}$ for any positive integer $k$.
\end{Lemma}

\Proof Suppose that \eqref{eq-7th} have two distinct solutions $x_1, x_2\in\mu_{q+1}$, then $x^7+tx^6+tx^4-x^3-x-t$ has a quadratic factor $x^2+ax+b$ satisfying $x_1+x_2=-a$ and $x_1x_2=b$ which implies $a^qb=a$. Moreover, the two solutions can be expressed as
\begin{eqnarray*}
 x_1=a-\sqrt{a^2-b},\; x_2=a+\sqrt{a^2-b}.
\end{eqnarray*}
Note that $x_1\ne x_2$ means $a^2\ne b$ which indicates that the case $b=-1$ in Lemma \ref{lem-7th-ab} cannot occur. Then, by Lemma \ref{lem-7th-ab} (2), we can obtain
\begin{eqnarray*}
 x_1=a-\sqrt{\theta}(b+1),\; x_2=a+\sqrt{\theta}(b+1),
\end{eqnarray*}
where $\theta^3-\theta-1=0$. Since $x^3-x-1$ is an irreducible polynomial over $\mathbb{F}_3$, then by \cite[Theorem 2.14]{Lidl-N} we have that $x^3-x-1=0$ has solutions in $\fqtwo$ if and only if  $2k\equiv 0\pmod{3}$, i.e., $k\equiv 0\pmod{3}$. Thus, if $k\not\equiv 0\pmod{3}$, then the second case in Lemma \ref{lem-7th-ab} cannot occur and then \eqref{eq-7th} cannot have two distinct solutions in $\mu_{q+1}$. Next we consider the case $k\equiv 0\pmod{3}$. In this case, we have $\theta\in \mathbb{F}_{3^3}\subseteq\fq$ due to  \cite[Theorem 2.14]{Lidl-N} .  Moreover, by $\theta^3-\theta-1=0$, we have $\theta^{13}=\theta\cdot\theta^3\cdot\theta^9=\theta(\theta+1)(\theta-1)=\theta^3-\theta=1$. This implies that $\theta^{(q-1)/2}=1$ since $q\equiv 1\pmod{26}$ if $k\equiv 0\pmod{3}$. On the other hand, by $x_1, x_2\in\mu_{q+1}$, we have $(a\pm\sqrt{\theta}(b+1))^{q+1}=1$ which leads to
  \begin{eqnarray*}
  a^q\sqrt{\theta}(b+1)+a\sqrt{\theta}^q(b^q+1)=0.
  \end{eqnarray*}
 This together with $a^q=a/b$ and $b^q=1/b$ gives $\sqrt{\theta}+\sqrt{\theta}^q=0$, a contradiction with $\theta^{(q-1)/2}=1$. Therefore, we can conclude that \eqref{eq-7th} cannot have two distinct solutions in $\mu_{q+1}$ for any positive integer $k$. This completes the proof. \done

Thus, we complete the proof of Conjecture \ref{conj-2} (2) according to Lemma \ref{lem-7th}.

To end this section, we point out that Conjecture \ref{conj-2} (1) can be discussed in the same way. To prove Conjecture \ref{conj-2} (1), we need to show that $\frac{-x^7+x^6+x}{x^6+x-1}=t$ has a unique solution for any $t\in\mu_{q+1}$. Similar to Lemma \ref{lem-7th-ab}, by a direct calculation we can show that $x^7+(t-1)x^6+(t-1)x-t$ has a quadratic factor $x^2+ax+b$, where $ab\ne 0$ and $a^qb=a$, only if $a, b$ satisfy $a^6+a^5b+a^5+a^4b-a^3b^2-a^3b-b^6-b^3-1=0$. Dividing $a^6$ on both sides gives $1+(b+1)/a+b/a^2-(b^2+b)/a^3=(b+1)^6/a^6-b^3/a^6$. Then, let $u=a^{-1}+a^{-q}=(b+1)/a$ and  $v=a^{-1}\cdot a^{-q}=b/a^2$ we can get $v^3-(u-1)v-(u^6-u-1)=0$, i.e., $(v-1)^3-(u-1)(v-1)-u^6=0$.
However, currently we do not know how to use this identity to prove that $x^2+ax+b=0$ cannot have distinct solutions in $\mu_{q+1}$ for even $k$.

\section{Conclusion remarks}\label{sec-cr}

In this paper, by analyzing the possible quadratic factors of a fifth-degree polynomial and a seventh-degree polynomial over $\mathbb{F}_{3^{2k}}$, two of the conjectures on permutation trinomials over $\mathbb{F}_{3^{2k}}$ proposed recently by Li, Qu, Li and Fu in \cite{LQLF} were settled. 

%

\section*{Acknowledgements}

This work was supported by the Norwegian Research Council.


\begin{thebibliography}{99}

\bibitem{BTT} C. Bracken, C.H.  Tan and Y. Tan, Binomial differentially 4 uniform permutations with high nonlinearity, Finite Fields Appl. 18(3)(2012), pp. 537-546.


 \bibitem{Carlitz-Wells} L.  Carlitz  and  C.  Wells, The  number  of solutions  of  a  special  system  of equations  in  a finite  field,  Acta  Arith. 12  (1966),  pp. 77-84.
 

 \bibitem{Dickson}L.E. Dickson, The analytic representation of substitutions on a power of a prime number of letters with a discussion of the linear group, Ann. of Math., 11 (1896), pp. 65-120.

 \bibitem{Dickson-58}  L.E.  Dickson, Linear   Groups  with  an  exposition  of  the  Galois  field  theory, Dover,  New York,  1958.

 \bibitem{Ding-QWYY} C. Ding, L. Qu, Q. Wang, J. Yuan and P. Yuan,  Permutation trinomials over finite fields with even characteristic, SIAM J. Dis. Math, 29 (2015), pp. 79-92.

 \bibitem{Hermite} Ch. Hermite, Sur les fonctions de sept lettres, C. R. Acad. Sci. Paris, 57 (1863), pp. 750-757.

 \bibitem{GS} Rohit Gupta and R.K. Sharma, Some new classes of permutation trinomials over finite fields with even characteristic, Finite Fields Appl. 41 (2016), pp. 89-96.

 \bibitem{Hou14} X. Hou, A class of permutation trinomials over finite fields, Acta Arith. 162 (2014), pp. 51-64.

  \bibitem{Hou} X. Hou, Permutation polynomials over finite fields--A survey of recent advances, Finite Fields Appl. 32 (2015), pp. 82-119.

 \bibitem{Hou15} X. Hou, Determination of a type of permutation trinomials over finite fields, II, Finite Fields Appl. 35 (2015), pp. 16-35.

  \bibitem{Li-Qu-Chen} K. Li, L. Qu and X. Chen, New classes of permutation binomials and permutation trinomials over finite fields, available online: http://arxiv.org/pdf/1508.07590.pdf.

  \bibitem{LQLF} K. Li, L. Qu, C. Li and S. Fu, New permutation trinomials constructed from fractional polynomials, available online: https://arxiv.org/pdf/1605.06216v1.pdf

  \bibitem{LH} N. Li and T. Helleseth,  Several classes of permutation trinomials from Niho exponents, submitted.
  
  \bibitem{LH2} N. Li and T. Helleseth, New permutation trinomials from Niho exponents over finite fields with even characteristic, available online: http://arxiv.org/pdf/1606.03768v1.pdf


 \bibitem{Lidl-N} R. Lidl and H. Niederreiter, Finite Fields, 2nd ed. Cambridge Univ. Press, Cambridge, 1997.

  \bibitem{Niederreiter-Robinson} N. Niederreiter and K.H. Robinson, Complete mappings of finite fields, J.  Austral.  Math. Soc. 33  (1982),  pp. 197-212.

 \bibitem{Park-Lee} Y.H. Park and J.B. Lee, Permutation polynomials and group permutation polynomials, Bull. Austral. Math. Soc. 63 (2001), pp. 67-74.

 \bibitem{Tu-Zeng-Hu} Z. Tu, X. Zeng and L. Hu, Several classes of complete permutation polynomials, Finite Fields and Appl., 25 (2014), pp. 182-193.
 
  \bibitem{Tu-Zeng-Hu-Li} Z. Tu, X. Zeng, L. Hu and C. Li, A class of binomial permutation polynomials, available online: http://arxiv.org/pdf/1310.0337v1.pdf.

  \bibitem{Tu-Zeng-J}  Z. Tu, X. Zeng and Y. Jiang, Two classes of permutation polynomials having the form $(x^{2^m}+x+\delta)^s+x$, Finite Fields Appl.  31(2015), pp. 12-24. 

 \bibitem{Tu-Zeng-L-H}  Z. Tu, X. Zeng, C. Li and T. Helleseth, Permutation polynomials of the form $(x^{p^m}-x+\delta)^s+L(x)$ over the finite field $\mathbb{F}_{p^{2m}}$ of odd characteristic, Finite Fields Appl.  34(2015), pp. 20-35. 

 \bibitem{Wan-Lidl} D. Wan and R. Lidl, Permutation polynomials of the form $x^rh(x^{(q-1)/d})$ and their group structure, Monatshefte. Math. 112 (1991), pp. 149-163.

 \bibitem{Zeng} X. Zeng, X. Zhu, and Lei Hu, Two new permutation polynomials with the form $(x^{2^k}+x+d)^s+x$ over $\mathbb{F}_{2^n}$. Appl. Algebra Eng. Commun. Comput. 21(2)(2010), pp. 145-150.

 \bibitem{ZTT} X. Zeng, S. Tian, and Z. Tu, Permutation polynomials from trace functions over finite fields, Finite Fields Appl. 35(2015), pp. 36-51.

\bibitem{ZZC} X. Zhu, X. Zeng, and Y. Chen, Some binomial and trinomial differentially 4-uniform permutation polynomials, Int. J. Found. Comput. Sci. 26(4)(2015), pp, 487-498.

 \bibitem{Zieve-subgroup} M. Zieve, Permutation polynomials on $\mathbb{F}_q$ induced form R\'{e}dei function bijections on subgroups of $\mathbb{F}_q^*$, available online: http://arxiv.org/pdf/1310.0776v2.pdf.

 \bibitem{Zieve-09} M. Zieve, On some permutation polynomials over $\mathbb{F}_q$ of the form $x^rh(x^{(q-1)/d})$, Proc. Amer. Math. Soc. 137 (2009), pp. 2209-2216.


\end{thebibliography}
\end{document}